\documentstyle[epsfig]{aprim10}
\input{epsf}
\newif\ifAMStwofonts
\ifoldfss
  \ifCUPmtlplainloaded \else
    \NewTextAlphabet{textbfit} {cmbxti10} {}
    \NewTextAlphabet{textbfss} {cmssbx10} {}
    \NewMathAlphabet{mathbfit} {cmbxti10} {} % for math mode
    \NewMathAlphabet{mathbfss} {cmssbx10} {} %  "   "    "
  \fi
  \ifAMStwofonts
    \ifCUPmtlplainloaded \else
      \NewSymbolFont{upmath} {eurm10}
      \NewSymbolFont{AMSa} {msam10}
      \NewMathSymbol{\upi}     {0}{upmath}{19}
      \NewMathSymbol{\umu}     {0}{upmath}{16}
      \NewMathSymbol{\upartial}{0}{upmath}{40}
      \NewMathSymbol{\leqslant}{3}{AMSa}{36}
      \NewMathSymbol{\geqslant}{3}{AMSa}{3E}

    \fi
  \fi
\fi % End of OFSS

\ifnfssone
  \newmathalphabet{\mathit}
  \addtoversion{normal}{\mathit}{cmr}{m}{it}
  \addtoversion{bold}{\mathit}{cmr}{bx}{it}
  \newmathalphabet{\mathbfit} % math mode version of \textbfit{..}
  \addtoversion{normal}{\mathbfit}{cmr}{bx}{it}
  \addtoversion{bold}{\mathbfit}{cmr}{bx}{it}
  \newmathalphabet{\mathbfss} % math mode version of \textbfss{..}
  \addtoversion{normal}{\mathbfss}{cmss}{bx}{n}
  \addtoversion{bold}{\mathbfss}{cmss}{bx}{n}
  \ifAMStwofonts
    \ifCUPmtlplainloaded \else
      \UseAMStwoboldmath
      \makeatletter
      \new@mathgroup\upmath@group
      \define@mathgroup\mv@normal\upmath@group{eur}{m}{n}
      \define@mathgroup\mv@bold\upmath@group{eur}{b}{n}
      \edef\UPM{\hexnumber\upmath@group}
      \new@mathgroup\amsa@group
      \define@mathgroup\mv@normal\amsa@group{msa}{m}{n}
      \define@mathgroup\mv@bold\amsa@group{msa}{m}{n}
      \edef\AMSa{\hexnumber\amsa@group}
      \makeatother
      \mathchardef\upi="0\UPM19
      \mathchardef\umu="0\UPM16
      \mathchardef\upartial="0\UPM40
      \mathchardef\leqslant="3\AMSa36
      \mathchardef\geqslant="3\AMSa3E
    \fi
  \fi
\fi % End of NFSS release 1

\ifnfsstwo
  \DeclareMathAlphabet{\mathbfit}{OT1}{cmr}{bx}{it}
  \SetMathAlphabet\mathbfit{bold}{OT1}{cmr}{bx}{it}
  \DeclareMathAlphabet{\mathbfss}{OT1}{cmss}{bx}{n}
  \SetMathAlphabet\mathbfss{bold}{OT1}{cmss}{bx}{n}
  \ifAMStwofonts
    \ifCUPmtlplainloaded \else
      \DeclareSymbolFont{UPM}{U}{eur}{m}{n}
      \SetSymbolFont{UPM}{bold}{U}{eur}{b}{n}
      \DeclareSymbolFont{AMSa}{U}{msa}{m}{n}
      \DeclareMathSymbol{\upi}{0}{UPM}{"19}
      \DeclareMathSymbol{\umu}{0}{UPM}{"16}
      \DeclareMathSymbol{\upartial}{0}{UPM}{"40}
      \DeclareMathSymbol{\leqslant}{3}{AMSa}{"36}
      \DeclareMathSymbol{\geqslant}{3}{AMSa}{"3E}
    \fi
  \fi
\fi % End of NFSS release 2

\ifCUPmtlplainloaded \else
  \ifAMStwofonts \else % If no AMS fonts
    \def\upi{\pi}
    \def\umu{\mu}
    \def\upartial{\partial}
  \fi
\fi

\title[Coronal Magnetic Field Measurements]{The Solar Coronal Magnetic Field Measurements With SOLARC}

\author[Liu and Lin]
       {Y. Liu$^{1,2}$ and H. Lin$^2$\\
        $^1$Yunnan Astronomical Observatory / National Astronomical Observatories,
Kunming, China\\
        $^2$Institute for Astronomy, University of Hawaii, Hawaii, USA}
\date{}

\pagerange{\pageref{firstpage}--\pageref{lastpage}}
\pubyear{2008}

\begin{document}

\maketitle

\label{firstpage}

\begin{abstract}
Direct solar coronal magnetic field measurements have become
possible since recent development of high-sensitivity infrared
detection technology. The SOLARC instrument installed on Mt.
Haleakala is such a polarimetric coronagraph that was designed for
routinely observing Stokes parameter profiles in near infrared (NIR)
wavelengthes. The Fe$^{+12}$ 1075 nm forbidden coronal emission line
(CEL) is potential for weak coronal magnetic field detection. As a
first step the potential field model has been used to compare with
the SOLARC observation in the Fe$^{+12}$ 1075 nm line \cite{liu08}.
It's found that the potential fields can be a zeroth-order proxy for
approaching the observed coronal field above a simple stable
sunspot. In this paper we further discuss several nodi that are
hampering the progress for reconstructing the real coronal magnetic
field structures. They include the well-known Van Vleck effect in
linear polarization signals, ignorance of the information of the NIR
emission sources (i.e., inversion problem of coronal magnetic
fields), a fat lot of global non-linear force-free field tools
available for better modeling coronal magnetic fields, and so on.

\end{abstract}


\section{Introduction}
Successful coronal magnetic field measurements by IR polarimetry
method have been reported recently \cite{lin00,lin04}. These first
coronal magnetic field maps offer the valuable opportunity to test
the various popular theoretical models for coronal magnetic field
structures.

Currently we have completed the study on the comparison between the
observation by SOLARC and the potential field model (Liu and Lin
2008; hereafter, paper I). We found that the observed linear and
circular polarization signals are well consistent with the simulated
results from the layers located just above the strong photospheric
field of a sunspot near the plane of the sky containing the solar
center. This conclusion is significant for both theoretic and
observational researches. {\it{First}}, the corona should be no
longer optically thin if large optical density of the coronal plasma
dominating somewhere in the path along the sight line, resulting the
use of a more reasonable integration function that is heavily
weighted toward particular layers close to sunspot than the
traditional uniform integration one. {\it{Second}}, since the
coronal NIR emission may be mainly contributed by a local coronal
region, then it is feasible to reveal its magnetic field structures
from the polarization signals observed. In another word, the
polarization signals are meaningful for tracing local coronal
magnetic field configurations. Obviously the coronal CEL
polarization data will provide rich new information and help improve
our understanding of the solar coronal physics. {\it{Third}}, it
demonstrates the  possibility for the quantitative comparison
between coronal magnetic field measurements and coronal models since
many early studies counted on the comparison only based on
extrapolated magnetic field configurations, i.e., House
\shortcite{house74}. The two-dimensional magnetic flux map by SOLARC
is an array of $16\times8$ by 128 optical fibers covering a
rectangle coronal region of about $0.30\times0.15$ solar radii$^2$
near the limb (Fig.~1). Furthermore, if each pixel of the spatially
resolved coronal magnetogram carries the emergent polarization
information for a single coronal emission source, then, {\it
fourth}, such a two-dimensional coronal magnetogram should be
mapping a real three-dimensional coronal fields and the
single-source inversion method to infer the magnetic field directly
from the polarimetric observation such as that proposed by Judge
\shortcite{judge07} can be confirmed. {\it{Fifth}}, following the
the first comparison between coronal magnetic field observation and
global potenial magnetic field model (paper I), linear and
non-linear force-free field methods will be used as the subsequence
to study their respective advantage and validity for modeling the
non-potential coronal magnetic fields.

Before using the coronal polarization signal measurements by the
promising IR CELs to clearly convey the information of the coronal
magnetic field structure in the near future, several difficulties,
briefly presented in the following sections, must be figured out or
partly improved.

%------------------------------------------------
\begin{figure} %Fig.~1
 \centerline{{\epsfxsize=8.5cm\epsffile{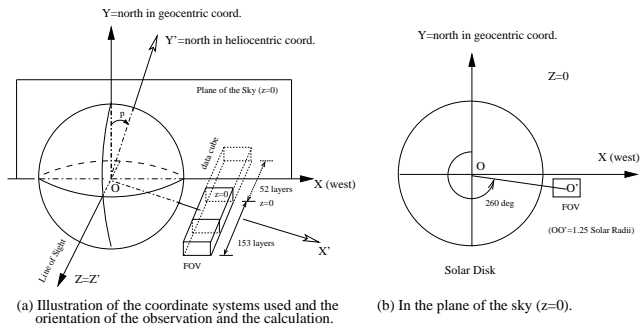}} }
 \caption[]{Showing location of the field of view (FOV) of the SOLARC observation on
 Apr 7, 2004, and the three-dimensional datacube calculated within it in the
 geocentric/heliocentric coordinate systems.  }
\end{figure}
%------------------------------------------------

\section{The Van Vlect effect}
Different from the transverse magnetic field observations for the
photosphere in which the judgement of the direction of the
transverse field vectors are subject to uncertainties due to the
famous $180\degr$ ambiguity in Zeeman effect, the transverse
magnetic field directions inferred from the linear polarization data
of the corona sustain not only the Zeeman effect ambiguity but the
$90\degr$ ambiguity due to the Van Vlect effect (Fig.~2, top frame).
House \shortcite{house74} had explained this phenomenon from the
point of the classical theory of electrodynamics. Fig.~2
demonstrates a simulation for the linear polarizations along a
coronal loop rooting at the solar limb. It is shown that both the
linear polarization amplitudes and directions are sensitive to the
location on the loop. However, the amplitude evolution is smooth
(Fig.~2, bottom frame), while the direction can change abruptly from
parallel to perpendicular relatively to the local tangent direction
of the loop. Thus the vector magnetic field directions (projected on
the plane of the sky) can not be resolved from one coronal
magnetogram. Theoretically, the reliability of three-dimensional
coronal vector tomographic techniques can be tested by synthesized
coronal Hanle and Zeeman effect observations to bring this issue to
a close \cite{kramar06}. However, obtaining long time-sequence
spectropolarimetric coronal observations with high quality is not
easy for ground-based instruments.

%----------------------------------------------------------------------
\begin{figure} %Fig.~2
 \centerline{{\epsfxsize=7.5cm\epsffile{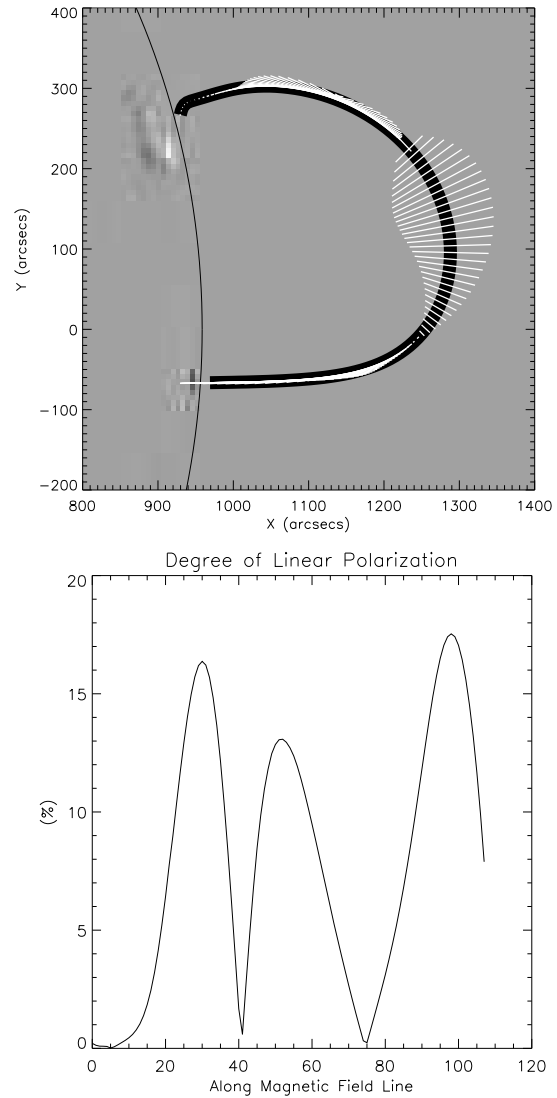}}}
 \caption[]{Illustration of the Van Vleck effect that plays an important role in modifying
 the linear polarizations along a coronal loop.
 {\it{Top}}: if the angle between the local magnetic field direction
 and the local solar radial direction is less than the Van Vleck angle
 ($54.7\degr$), the linear polarization direction (white bar) will keep
 parallel to the magnetic field vector at the scattering center;
 otherwise, it will become perpendicular (e.g., those close to the loop's top).
 Moreover, when this angle is close to 54.7$\degr$ (i.e., $asin^{-1}\sqrt{2/3}$), the intensity of the
 linear polarization will be approaching to zero. {\it{Bottom}}: the linear
 polarization degree evolution along the coronal loop. }
\end{figure}
%----------------------------------------------------------------------

\section{Global non-linear force-free magnetic field models}
We need model coronal magnetic fields in spherical geometry for the
purpose of direct comparison with the large-scale SOLARC
observations covering coronal regions close to the solar limb.
Although there are some non-linear force-free field methods that do
not depend on the use of a specific coordinate system, most of the
codes available are implemented in Cartesian geometry, ignoring the
more general application in the case of the spherical coordinate
system.A recent practical tool using simple reference functions,
formally similar to the Green's function, has been developed and
improved \cite{yan06,he08}, which will be utilized for the
force-free models testing in our future work. This method is
convenient in the direct photospheric boundary integration for the
magnetic field component calculations for any interested coronal
point. It is noted that Wiegelmann \shortcite{wie07} presented a new
code for the extrapolation of non-linear force-free field in
spherical coordinates, but this method is not applied and still
under development.

%------------------------------------------------
\begin{figure} %Fig.~3
 \centerline{{\epsfxsize=8.5cm\epsffile{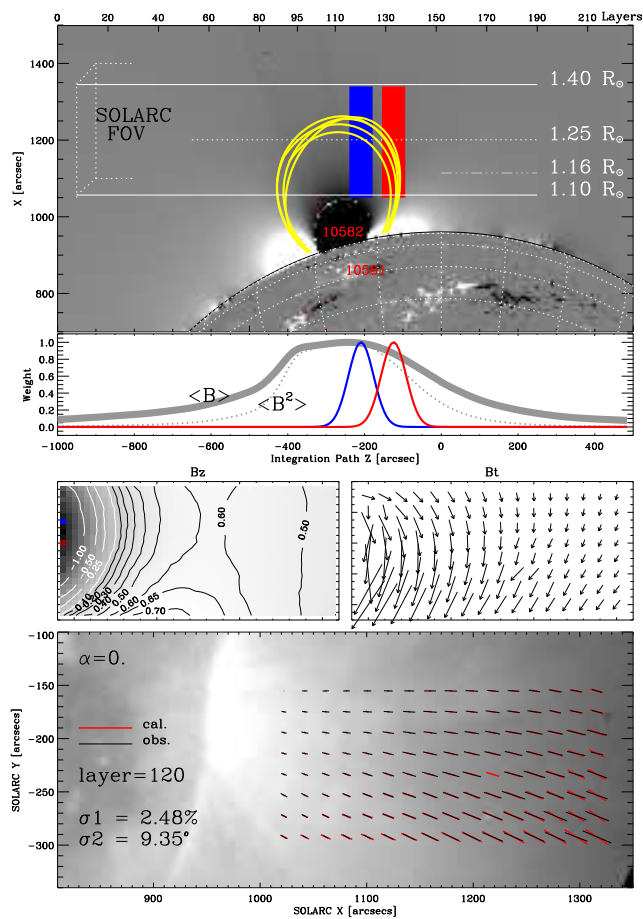}} }
 \caption[]{Demonstration for SOLARC observation and synthesized result. The coordinate system
 is the same as in Figure 2 and in the paper (Liu and Lin 2008). The two frames in the
 {\it{top two rows}}: illustrate the assumed emission sources and Stokes parameter integration
 Gaussian functions, in which the line in blue represents the Gaussian function profile centered
 at layer 120, while the one in red, centered at layer 130 just above the sunspot of
 AR 10582. The two frames in the {\it{third row}}: for the line-of-sight magnetic
 flux map at layer 130 (note, negative/positive fluxes are shown with white/black )
 and the transverse magnetic field vectors at layer 120, respectively. {\it{Bottom row}}: comparison
 between observed and synthesized linear polarizations for layer 120 that are overlaid on an EUV
 image.}
\end{figure}
%----------------------------------------------------------------------

\section{Single-source inversion}
The lack of knowledge of source regions of the IR coronal radiation,
including the coronal density and temperature distributions, is the
greatest uncertainty in our study with the SOLARC measurements.
Nevertheless, based on decades of observations, it is well known
that strong coronal emissions are always associated with active
regions. This experience was confirmed by the study of the careful
quantitative comparison between SOLARC data and potential field
extrapolation (paper I). Both circular and linear polarization
signals simulated with designed weighting functions (in red and
blue, respectively) are shown in Fig.~3. The vectors of the
transverse magnetic fields, synthesized within the SOLARC field of
view for the layer close to the sunspot AR 10582, are not parallel
to the directions of the linear polarization for most pixels! The
linear polarization directions seem keep open at the height ranging
from 0.1 to 0.4 solar radii, while the magnetic fields are closed.
The main reason for the close of the magnetic field lines is due to
the direct magnetic field connection between the sunspot AR 10582
and 10581 that form a large bright EUV coronal loops system in the
north-south direction obviously above the west limb (Fig.~2 in Lin
et al. 2004). It's suggested that CEL radiation may originate from a
region close to the strongest photospheric magnetic feature in the
active region with a small spatial scale comparable to the
characteristic size of the coronal loops seen in the intensity
images (Paper I). Triggered by this thought, we make a tentative
examination for all the five coronal regions observed by SOLARC. The
result is shown in Fig.~4. Note that the parameters used in the
calculation are the same as in paper I. Unexpectedly, one-valley
feature in the $\sigma{_{LP}}$ profiles is evident for most pixels
in the five coronal regions. However, this feature may be ruined by
the Van Vleck effect as seen in the first five or six columns in the
first two rows. At the higher heigh, the $\sigma{_{LP}}$ profiles
are more obviously featured by one valley. Does the the one-valley
feature indicate the strongest radiation CEL source location? It is
a critical question related to the coronal magnetic field inversion.

\section{conclusion}
We list and discuss some key problems in the probing of coronal
magnetic field in the short review. The difficulties are expected to
be improved in virtue of the recent launched {\it STEREO} mission
and the proposed vector coronal tomography techniques.

%------------------------------------------------
\begin{figure} %Fig.~4
 \centerline{{\epsfxsize=8.5cm\epsffile{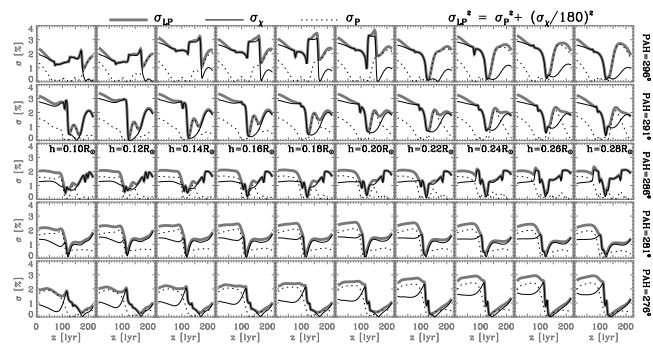}}
 }
 \caption[]{Inversion for NIR emission sources based on the $\sigma_{LP}$
 profiles. The results support the single-source opinion for CEL radiation for some cases.
 The parameters marked in the figure are the same as in paper I.}
\end{figure}
%----------------------------------------------------------------------

\label{lastpage}

\clearpage

\end{document}